\documentclass[conference]{IEEEtran}
\IEEEoverridecommandlockouts
\usepackage{cite}
\usepackage{amsmath,amssymb,amsfonts}
\usepackage{algorithmic}
\usepackage{graphicx}
\usepackage{textcomp}
\usepackage{float}
\usepackage{xcolor}
\usepackage{enumitem}
\usepackage{xurl}
\usepackage{hyperref} 
\def\BibTeX{{\rm B\kern-.05em{\sc i\kern-.025em b}\kern-.08em
    T\kern-.1667em\lower.7ex\hbox{E}\kern-.125emX}}
\newcommand{\ra}[1]{\renewcommand{\arraystretch}{#1}}
\usepackage{booktabs}
\usepackage{makecell}
\begin{document}

\title{Multi-Agent Reinforcement Learning in Cybersecurity: From Fundamentals to Applications
\thanks{This research is supported by armasuisse Science and Technology.}
}

\author{\IEEEauthorblockN{Christoph R. Landolt}
\IEEEauthorblockA{\textit{Cyber-Defence Campus} \\
\textit{\& Eastern Switzerland University of Applied Sciences}\\
Rapperswil, Switzerland \\
\url{https://orcid.org/0009-0002-7031-3291}}
\and
\IEEEauthorblockN{Christoph Würsch}
\IEEEauthorblockA{\textit{Institute for Computational Engineering} \\
\textit{Eastern Switzerland University of Applied Sciences}\\
Rapperswil, Switzerland \\
\url{https://orcid.org/0000-0002-1337-3477}}
\and
\IEEEauthorblockN{Roland Meier}
\IEEEauthorblockA{\textit{Cyber-Defence Campus} \\
\textit{armasuisse Science and Technology}\\
Thun, Switzerland \\
\url{https://orcid.org/0000-0002-8268-9037}}
\and
\IEEEauthorblockN{Alain Mermoud}
\IEEEauthorblockA{\textit{Cyber-Defence Campus} \\
\textit{armasuisse Science and Technology}\\
Thun, Switzerland \\
\url{https://orcid.org/0000-0001-6471-772X}}
\and
\IEEEauthorblockN{Julian Jang-Jaccard}
\IEEEauthorblockA{\textit{Cyber-Defence Campus} \\
\textit{armasuisse Science and Technology}\\
Thun, Switzerland \\
\url{https://orcid.org/0000-0002-1002-057X}}
}

\maketitle

\begin{abstract}
Multi-Agent Reinforcement Learning (MARL) has shown great potential as an adaptive solution for addressing modern cybersecurity challenges. MARL enables decentralized, adaptive, and collaborative defense strategies and provides an automated mechanism to combat dynamic, coordinated, and sophisticated threats. This survey investigates the current state of research in MARL applications for automated cyber defense (ACD), focusing on intruder detection and lateral movement containment. Additionally, it examines the role of Autonomous Intelligent Cyber-defense Agents (AICA) and Cyber Gyms in training and validating MARL agents. Finally, the paper outlines existing challenges, such as scalability and adversarial robustness, and proposes future research directions. This also discusses how MARL integrates in AICA to provide adaptive, scalable, and dynamic solutions to counter the increasingly sophisticated landscape of cyber threats. It highlights the transformative potential of MARL in areas like intrusion detection and lateral movement containment, and underscores the value of Cyber Gyms for training and validation of AICA.\\
\\
\textit{This paper was originally presented at the NATO Science and Technology Organization Symposium (ICMCIS) organized by the Information Systems Technology (IST) Panel, IST-209-RSY – the ICMCIS, held in Oeiras, Portugal, 13-14 May 2025}\\
\end{abstract}


\section{Introduction}
\label{introduction}
The World Economic Forum estimates that cybercrime will cost the global economy \$10.5 billion in 2025 \cite{muggah2023global}. These costs are influenced by the rapid evolution of cyber threats, which often challenge conventional defense mechanisms. Traditional approaches to cybersecurity, which are often static and reactive, might struggle to effectively counter dynamic and sophisticated threats \cite{cyberinsight_dynamic_static}. As a result, adaptive and proactive methods show great potential to respond to these developments and improve resilience against emerging risks \cite{kenyon2021traditional_security, kudelski2024threat_landscape}.

To this end, DARPA launched the Cyber Grand Challenge \cite{darpa2016cybergrandchallenge} to develop autonomous systems that can detect vulnerabilities and apply patches autonomously. This initiative culminated with systems such as Mayhem \cite{8328972}, Xandra \cite{8328984}, and Mechanical Phish \cite{8328966} demonstrating autonomous defense at machine speed by identifying and patching vulnerabilities in real-time \cite{darpa2016cybergrandchallenge}. However, while these systems could react quickly to known threats, they could not adapt independently to new, unforeseen threats \cite{208121}. 

To detect such threats, Intrusion Detection Systems (IDSes) typically employ in addition to static \emph{rule-based methods}, which rely on predefined signatures or patterns to identify known vulnerabilities, more dynamic \emph{anomaly-based approaches}, which use behavioral baselines and machine learning to detect deviations indicative of previously unknown vulnerabilities \cite{bsi_2022}. In addition, techniques such as fuzzing, where systems are tested with unexpected or corrupted inputs to identify bugs or crashes, have uncovered known and unknown vulnerabilities. These approaches help improve cybersecurity by automating the detection of vulnerabilities and enabling automated defense \cite{9166552}.

Building on these advances, Reinforcement Learning (RL) presents a promising framework for developing more dynamic and adaptive systems. RL can overcome the limitations of traditional approaches by enabling agents to learn independently through \emph{trial and error} and adapt to evolving attack strategies \cite{Uprety_2021, ADAWADKAR2022105116}. This dynamic adaptation is critical for detecting new attack strategies and \emph{attack graphs} that rule-based systems cannot identify \cite{GUO2023175, Hu2019}. RL's continuous learning capabilities enable it to adapt to new attack patterns over time, making it more robust and able to defend against  unknown threats \cite{cengiz2023reinforcement, Hu2019, GUO2023175}.

Real-world scenarios are often dynamic, as attackers conduct coordinated attacks while defenders must protect distributed systems; these dynamics are challenging to model and train effectively with a single-agent approach. Multi-Agent Reinforcement Learning (MARL), in contrast, offers collaborative and adversarial interactions between agents, enabling realistic training and adaptive strategies in complex, distributed environments \cite{kunz2023multiagentcyberbattlesimrlcyber}. MARL has shown great promise for enabling adaptive and collaborative defense strategies \cite{vyas2023automatedcyberdefencereview, palmer2024deepreinforcementlearningautonomous, oesch2024pathautonomouscyberdefense}.

This survey addresses the need for innovative approaches to modern cybersecurity challenges, focusing on how MARL and the development of AICA can provide adaptive, scalable, and dynamic solutions to counter the increasingly sophisticated landscape of cyber threats. It focuses on key areas such as moving target defense, and lateral movement containment because these areas are critical for detecting threats, enhancing system resilience, and countering sophisticated attacks like advanced persistent threats, where MARL can have the greatest impact. Unlike prior reviews, this paper emphasizes MARL's ability to handle multi-agent dynamics in adversarial settings. It focuses on its promising use cases for the development of AICA \cite{kott2023autonomousintelligentcyberdefenseagent}, which are adaptive, self-learning systems designed to detect, mitigate, and respond to cyber threats in real-time within ACD applications.

This paper is structured as follows: Section II reviews related surveys and motivates this paper, Section III discusses multi-agent interactions and MARL models, Section IV explores training environments for the training of AICAs, Section V highlights practical applications of MARL in cybersecurity, and Section VI concludes the state of research in the field.

\section{Related Surveys and Motivation}
\label{related_surveys}

The literature on MARL covers various research topics, with general reviews and books such as \cite{gronauer2022multi, huh2024multiagentreinforcementlearningcomprehensive, marl-book}. However, the application of reinforcement learning in cybersecurity is still relatively new, and as such, only a few review papers specifically address the current state of this research. Notable reviews that explore the use of MARL in cybersecurity include those by S. Oesch et al. \cite{oesch2024pathautonomouscyberdefense}, S. Vyas et al. \cite{vyas2023automatedcyberdefencereview}, G. Palmer et al. \cite{palmer2024deepreinforcementlearningautonomous}, and S. Finistrella et al. \cite{finistrella2024multi}.

In contrast to these existing reviews, our paper covers the foundations of MARL and investigates its potential for decentralized, collaborative defense strategies, particularly in dynamic and adversarial scenarios. As computer networks continue to evolve and grow increasingly complex, the challenges in cybersecurity are intensifying. Recent research highlights the role of evolutionary game methods in enhancing defensive capabilities through internal decision-making and learning mechanisms \cite{9394344}. Building on this foundation, we survey the advantages of collaborative decision-making in multi-agent systems, emphasizing coordinated interactions among defense agents using MARL. While other surveys focus primarily on specific RL techniques for specific cybersecurity applications, our work focuses on the foundations of MARL frameworks and their potential to handle complex, multi-agent interactions such as intrusion detection and lateral movement containment.

Additionally, we explore training environments for dynamic simulations and testing adaptive strategies to bridge the gap between simulation and practical use of AICAs.

\subsection{Motivation of MARL for Cyber Defense}
\label{limitations_sarl}

RL has shown potential in cyber defense due to its ability to discover new strategies through \emph{trial and error} without relying on new data. However, the applicability of Single-Agent Reinforcement Learning (SARL) in large and dynamic environments such as Cyber-Physical Systems (CPS) and computer networks has some limitations. This section discusses these limitations and how MARL overcomes them.

\subsubsection{Lack of Scalability}
In SARL, an agent controls an entire environment and learns from its interaction with it. As described in Section \ref{introduction} (Introduction), this approach has scalability issues in complex environments such as distributed systems, where decentralized and real-time decisions are necessary. Studies on state abstractions \cite{abel2019state_abstraction, abel2016state_abstraction} show that the computational complexity of SARL increases exponentially with larger state and action spaces, making learning impractical in such large systems.

MARL addresses this issue by providing a framework to train distributed agents, which can learn and act locally while pursuing system-wide goals. As described in Section \ref{games_of_ma_interactions}, these agents' training is often formulated as a partially observable stochastic game (POSG), where agents act under partial observability and receive rewards based on their actions. A specific subclass of POSGs, decentralized partially observable Markov decision processes (Dec-POMDPs), features decentralized policies and often considers joint rewards \cite{marl-book}. In MARL, Dec-POMDPs enable agents to act independently and collaboratively. However, addressing scalability remains an active area of research and often necessitates approximations or abstractions like factored representations, shared policies, or hierarchical approaches \cite{abel2016state_abstraction}.

\subsubsection{Inability to Handle Multi-Agent Dynamics} Multiple entities (attackers, defenders, neutral agents) interact within the same shared environment in cybersecurity scenarios. This challenge complicates the causality between action and observation in SARL, as SARL cannot effectively learn the different objectives and interactions between multiple agents. Research in adversarial learning \cite{abel2019state_abstraction_compression} demonstrates that single-agent approaches fail to generalize effectively when multiple opposing agents dynamically change their strategies due to the non-stationarity of the learning problem in multi-agent settings.
To address this, MARL uses frameworks like Nash Equilibria \cite{nash1951non_cooperative_games} to provide a theoretical basis for optimizing strategies in adversarial settings. Further methods like value decomposition networks \cite{sunehag2018value_decomposition} are effective in cooperative settings.

\subsubsection{Coordination Challenges} 
In distributed systems such as computer networks, cooperative behavior (exchange of information, communication) across different subsystems is often required. Tasks such as simultaneous threat detection and resource allocation cannot be efficiently accomplished by SARL \cite{abel2018state_abstractions}. MARL facilitates such coordination and collaboration challenges by enabling information exchange between agents and often leveraging global rewards. Techniques such as value decomposition \cite{sunehag2018value_decomposition} and state-action abstractions \cite{abel2020value_preserving_state_action} provide tools for efficient collaboration, leading to improved robustness and performance in applications such as cyber defense.

\section{Models of Multi-Agent Interaction}
\label{models_of_ma_interactions}
A Multi-Agent System (MAS) consists of agents that coexist in a shared environment to achieve individual or collective goals. These agents observe their environment, communicate with each other, and perform actions autonomously. Depending on the scenario, agents may compete for resources or collaborate to achieve shared objectives. Constraints such as partial observability—where each agent only observes a portion of the environment—and limited communication bandwidth significantly influence their strategies and behaviors \cite{huh2024multiagentreinforcementlearningcomprehensive}.
Agent-environment interactions follow a closed-loop dynamic: agents take actions, the environment transitions states, and feedback is provided as reward signals. Fig. \ref{fig:rl_control_loop} visualizes such a closed-loop RL system with one agent:
    \begin{figure}[H]
        \centering
        \includegraphics[width=0.4\textwidth, trim=0cm 0.3cm 0cm 0.3cm, clip]{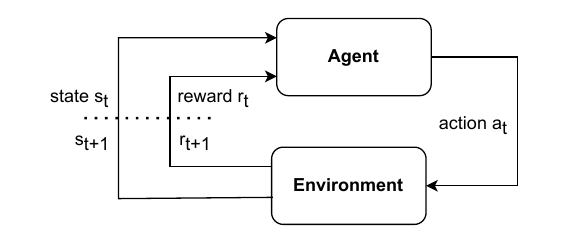} 
        \caption{Reinforcement Learning (RL) control loop, inspired by Sutton et al. \cite{Sutton2005ReinforcementLA}. The figure illustrates the interaction between a \textit{RL-Agent} and its \textit{environment} over discrete time steps. At each step $t$, the agent observes the current state $s_t$ and reward $r_t$ from the environment, selects an action $a_t$, and sends it to the environment. The environment then transitions to a new state $s_{t+1}$ and provides a new reward $r_{t+1}$, completing the loop.}
        \label{fig:rl_control_loop}
    \end{figure}

These iterative processes drive adaptation and learning, forming the foundation for mathematical models of agent dynamics discussed in the following sections \cite{shoham2008multiagent}.

\subsection{Game-Based Models}
\label{games_of_ma_interactions}
\begin{figure}[H]
        \centering
        \includegraphics[width=0.4\textwidth, trim=0cm 0.3cm 0cm 0.3cm, clip]{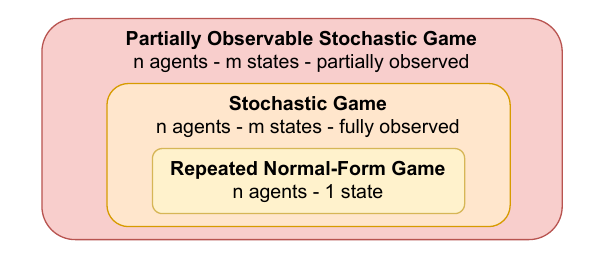} 
        \caption{Nested structure of game models. The \textit{Partially Observable Stochastic Game} includes $n$ agents and $m$ partially observable states. A simpler model is the \textit{Stochastic Game}, which assumes the complete observability of the states, while the \textit{Repeated Normal-Form Game} involves $n$ agents interacting in a single state. Adapted from \cite{marl-book}.}
        \label{fig:hierarchy_of_game_models}
    \end{figure}

The formal models and frameworks of interactions in multi-agent environments are grounded in game theory, providing the foundation for understanding and designing these systems and RL paradigms. S. Albrecht et al. \cite{marl-book} describe a hierarchy of these models, which is illustrated in Fig. \ref{fig:hierarchy_of_game_models}:
\setlist[enumerate,1]{leftmargin=*} 
\begin{enumerate}
    \item \emph{Repeated Normal-Form Games}  extend normal-form games, representing single interactions between agents, by repeating these interactions sequentially over a finite or infinite horizon. In cybersecurity, these games model the interactions between attackers and defenders, such as allocating resources (e.g., honeypots) to defend critical network components against evolving attack strategies over time \cite{1273013}. By identifying equilibrium points, these models enable robust and optimal defense mechanisms in dynamic threat environments \cite{10.5555/1558013.1558108}.
    % \vspace{0.25em}
    \item \emph{Stochastic Games} extend Markov Decision Processes (MDPs) to multi-agent settings, where the joint actions of multiple agents influence both the transition dynamics and the rewards. Agents aim to optimize strategies while accounting for these interdependencies. For example, in cybersecurity, stochastic games can also be used to model some more sophisticated interactions between attackers and defenders, where the probability of transitioning to a new system state (e.g., a compromised or secure state) depends on the joint actions of both parties, such as an attacker choosing specific exploitation strategy and a defender choosing a response (e.g., blocking an IP-address) \cite{g15040028}. Another application is securing industrial control systems, where stochastic games capture the uncertainty in adversarial actions and environmental factors, allowing defenders to predict the likelihood of system breaches and proactively allocate resources to minimize risks \cite{Etesami2019DynamicGI}.
    % \vspace{0.25em}
    \item \emph{Partially Observable Markov Decision Processes (POMDPs)} address situations where agents have limited visibility of the environment and must rely on partial observations to make decisions. In cybersecurity, POMDPs can model intrusion detection systems where defenders must make decisions based on incomplete and noisy data about potential threats in a network \cite{8325528}.
    % \vspace{0.25em}
    \item \emph{Decentralized POMDPs (Dec-POMDPs)}  is an advanced extension of POMDPs and involves multiple agents operating under partial observability without centralized coordination. They base decisions on local observations while receiving a shared reward.\\
For example, in cybersecurity, Dec-POMDPs can be used to model the collaboration of distributed IDSes, where agents monitor different parts of a network and must collaboratively identify and respond to attacks based on their local observations \cite{Yu_2018}.
\end{enumerate}

These game-theoretic frameworks allow modelling complex inter-agent dynamics and address uncertainty and partial observability in centralized and decentralized scenarios.

MARL builds on these principles by including learning-based methods to find an equilibria solution. These methods allow agents to approximate optimal strategies and adapt to dynamic environments. By leveraging approximation techniques, MARL provides a scalable approach to finding solutions in high-dimensional, multi-player systems where the direct computation of equilibria is computationally expensive and often infeasible \cite{marl-book}.

% \newpage
\subsection{Reinforcement Learning-Based Interactions}
 MARL is a particular use case of RL that investigates and models such multi-agent dynamics. As in RL, the agents are motivated by maximizing their rewards. By designing this reward, the self-interested agents can pursue mutual or opposite interests by competing. This section explores how the RL paradigm can be extended to multiple agents. This extension enables the development of decentralized systems or modeling interactions among multiple actors.

\subsubsection{Learning in a Multi-agent Environment}
\label{marl_introduction}
\begin{figure}[H]
        \centering
        \includegraphics[width=0.5\textwidth, trim=0cm 0.1cm 0cm 0cm, clip]{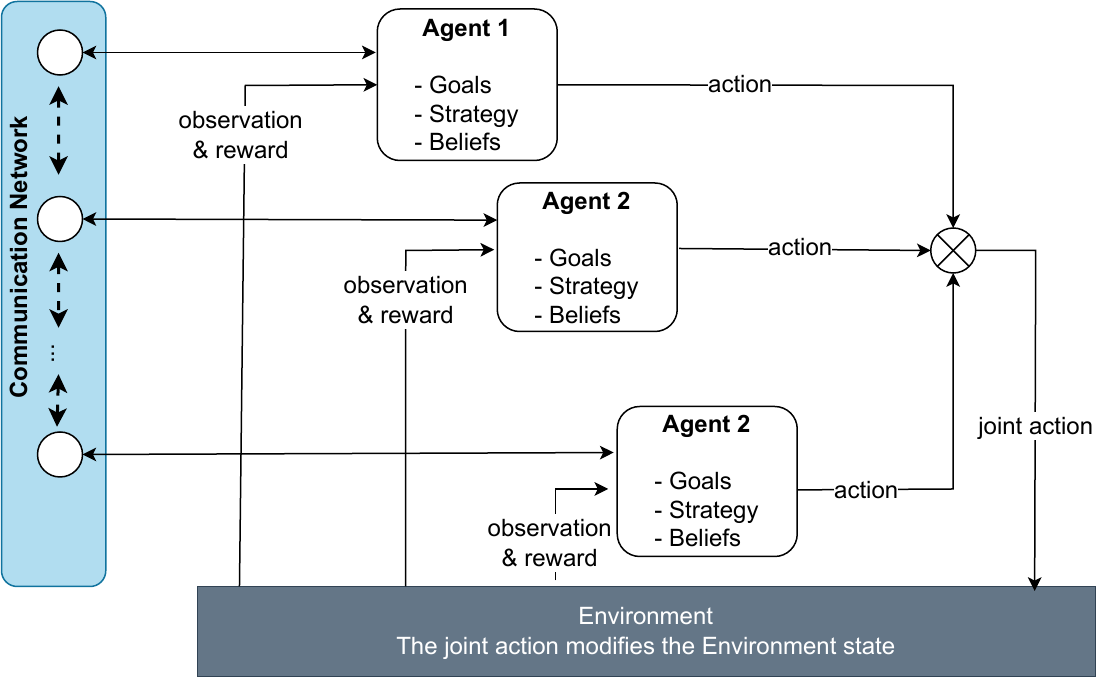} 
        \caption{Interaction between multiple agents, an optional communication network, and the environment. Each agent can have its own goals, strategies, and beliefs. The agent receives individual observations and rewards from the environment. The actions of all agents are combined into a joint action that influences the state of the environment. This joint action introduces complexity in determining the causal relationship between an individual agent's actions and the resulting rewards and observations.}
        \label{fig:multi-agent_reinforcement_learning}
    \end{figure}

MARL uses algorithms to learn optimal policies for a set of agents in a MAS \cite{marl-book}. Similar to SARL, the policies are learned via \emph{trial and error} to maximize the agents' cumulative rewards. Unlike pure trial and error, MARL incorporates informed exploration based on previously learned policies, addressing the exploration-exploitation dilemma by balancing the discovery of new actions with leveraging those currently believed to be optimal. While exploration can uncover better actions, it may also  result in lower rewards during the learning process. Fig. \ref{fig:multi-agent_reinforcement_learning} shows a basic schematic of the MARL training loop.

\paragraph{MARL Training Paradigms}
In MARL scenarios, agents operate in a game-theoretic framework, where their actions and strategies influence each other. An optimum of the agent's interplay is achieved by reaching an equilibrium point—a state in which no agent is incentivized to deviate from its chosen strategy. The pursuit of such equilibria is guided by three fundamental training schemes: cooperative, competitive, and mixed-interests:
\setlist[itemize,1]{leftmargin=*} 
\begin{itemize}
    \item \textbf{Cooperative:} Cooperative MARL focuses on agents learning to work together to achieve a common goal while maximizing a shared reward. The reward structure is designed to reinforce collective performance, as illustrated in Fig. \ref{fig:cooperative_game}, such as distributed intrusion detection \cite{tampuu2015multiagentcooperationcompetitiondeep}.
    \begin{figure}[H]
        \centering
        \includegraphics[width=0.5\textwidth, trim=0cm 0.05cm 0cm 0.3cm, clip]{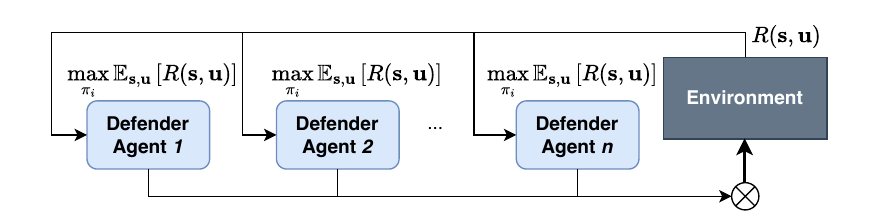} 
        \caption{Cooperative Defense of a System by $n$ Distributed Defenders through a cooperative MARL-System, where the defender agents collaborate to maximize their expected reward $\mathbb{E}_{s, u}[R(s, u)]$, with $R(s, u)$ denoting the reward received when all agents take a joint action $u$ in state $s$. Each defender agent operates independently, optimizing its policy $\pi_i$ based on local observations and rewards.}
        \label{fig:cooperative_game}
    \end{figure}
    This collaborative process can described through a mathematical framework that expresses the joint actions and policies of agents in a cooperative setting, as outlined below:
    \begin{itemize}[label=-]
        \item Let $\mathbf{u} = (u_1, u_2, \ldots, u_N)$ represent the joint action of all agents, where $u_i$ is the action taken by agent $i$. Actions may be executed either in parallel or sequentially. In sequential systems, each agent acts one at a time, often determined by a predefined order or environmental rules, whereas, in parallel systems, all agents act simultaneously, with potential tie-breaking mechanisms applied to resolve conflicting actions \cite{terry2021pettingzoogymmultiagentreinforcement}.
        \item  Let $\mathbf{s}$ represent the state of the environment.
        \item Let $\pi_i(u_i \vert s)$ be the policy of agent $i$, defining the probability of taking action $u_i$ given state $s$.
        \item Let $R(\mathbf{s}, \mathbf{u})$ be the collective reward received from the environment based on the state $\mathbf{s}$ and the joint action $\mathbf{u}$. The collective reward can be expressed as the sum of individual rewards, where each agent $i$ receives a reward based on the global state $\mathbf{s}$ and its own action $u_i$: $R(\mathbf{s}, \mathbf{u}) = \sum_{i=1}^N R_i(\mathbf{s}, u_i)$, where $N$ is the number of agents, and $R_i(\mathbf{s}, u_i)$ is the reward received by agent $i$ according to its policy $\pi_i(u_i \vert \mathbf{s})$.
    \end{itemize}
    Each agent aims to contribute to maximizing the collective reward by learning an optimal policy $\pi_i^*(u_i \vert s)$. The optimization of the collective reward results in the global optimization objective $\max_{\{\pi_i\}_{i=1}^N} \mathbb{E}[R(s, u(\pi_i))]$, where $\mathbb{E}[R(s, u)]$ denotes the expected reward given state $s$ and joint action $u$. The learning process of each agent contributes to finding a Nash equilibrium or a Pareto-efficient set of strategies that ensure the collective maximization of $R(\mathbf{s}, \mathbf{u})$.
    % \vspace{1em}
    \item \textbf{Competitive:} In competitive scenarios, agents try to maximize their own rewards while minimizing their opponents (zero-sum games), e.g., chess and Go, or in attacker-defender scenarios in cybersecurity \cite{tampuu2015multiagentcooperationcompetitiondeep}.\\
Fig. \ref{fig:Zero-Sum} visualizes a basic schematic of the MARL training loop in an adversarial setup.
    \begin{figure}
        \centering
        \includegraphics[width=0.4\textwidth, trim=0cm 0.5cm 0cm 0.6cm, clip]{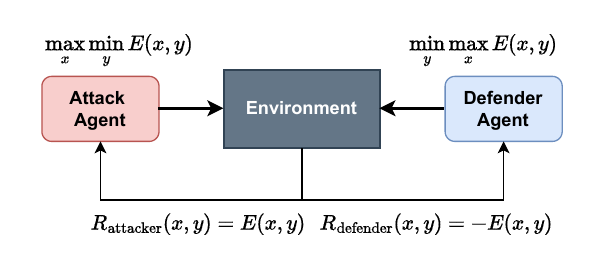} 
        \caption{Zero-sum interaction between an attacker and defender agent in an environment. The attack agent aims to maximize its reward $R_{\text{attacker}}(x, y) = E(x, y)$ by selecting actions $x$. In contrast, the Defender Agent aims to minimize its reward $R_{\text{defender}}(x, y) = -E(x, y)$ by choosing actions $y$.}
        \label{fig:Zero-Sum}
    \end{figure}
    In a zero-sum attacker-defender game, the attacker chooses actions $x$ from a set $A$, and the defender chooses actions $y$ from a set $D$, with the payoff matrix $M[i][j]$ representing the attacker's reward when action $i$ is met by the defender's $j$. The defender's payoff is $-M[i][j]$. Mixed strategies $x$ (attacker) and $y$ (defender) represent probabilistic choices, summing to $1$. The expected payoff for the attacker is $E(x, y) = x^T M y$, with the defender minimizing $E(x, y)$. The game’s solution is a Nash equilibrium, satisfying $\max_x \min_y E(x, y) = \min_y \max_x E(x, y)$.
    % \vspace{1em}
    \item \textbf{Mixed-Interest:} Agents engage in cooperative and competitive dynamics where they might have partially aligned and partially conflicting goals, as illustrated in Fig. \ref{fig:mixed_interest}. This situation usually shows up in trading, traffic, and multi-player video games \cite{lowe2020multiagentactorcriticmixedcooperativecompetitive}.
    \begin{figure}[H]
        \centering
        \includegraphics[width=0.45\textwidth, trim=0cm 1.0cm 0cm 1.8cm, clip]{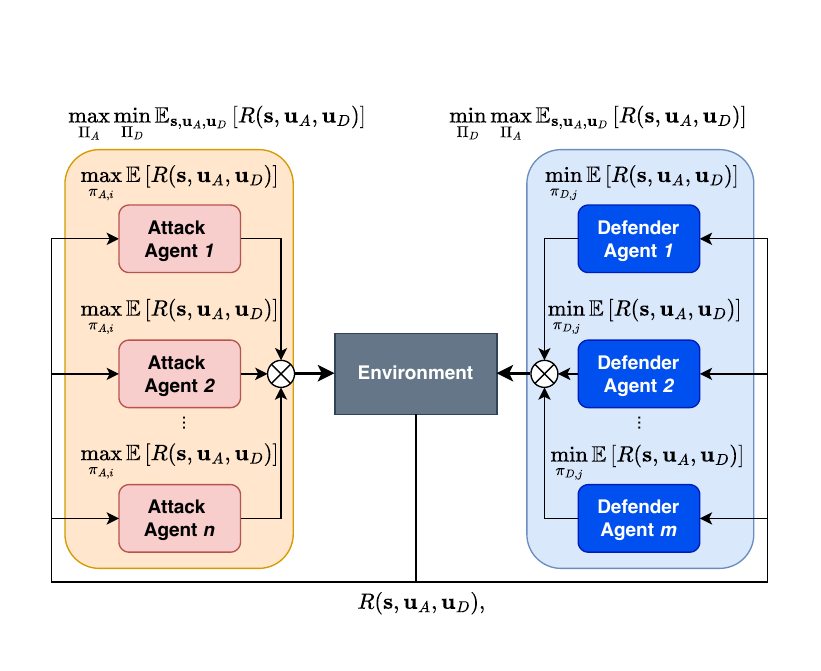} 
        \caption{Adversarial setup where $n$ attack agents maximize their expected reward $\mathbb{E}[R(s, \mathbf{u_A}, \mathbf{u_D})]$, while $m$ defender agents minimize it. Both groups optimize their respective policies ($\pi_{A,i}$ and $\pi_{D,j}$) and interact through a shared environment, with joint actions $\mathbf{u_A}$ and $\mathbf{u_D}$ defining state transitions and rewards.}
        \label{fig:mixed_interest}
    \end{figure}
    Each attacker $i$ optimizes its policy $\pi_{A,i}(u_{A,i}|s)$ to maximize the collective reward for the attackers, aiming to achieve a coordinated strategy represented by the joint attacker policy $\Pi_A(u_A|s)$. Conversely, each defender $j$ optimizes its policy $\pi_{D,j}(u_{D,j}|s)$ to minimize the attackers’ reward (equivalently, maximize their reward), with their joint policy represented by $\Pi_D(u_D|s)$.
    
    Globally, the problem is formulated as a zero-sum optimization challenge where the attackers and defenders collectively aim to find a Nash equilibrium, which is achieved when the attackers maximize their global reward $J_A(\Pi_A, \Pi_D)$, while the defenders minimize it, ensuring that neither group can unilaterally improve their outcome. The equilibrium defines the value $v$ of the game, where the attackers’ global reward equals $v$ and the defenders’ global reward equals $-v$, ensuring the zero-sum property. The collective optimization is achieved through iterative policy updates and RL techniques to converge toward these equilibrium policies $\Pi_A^*$ and $\Pi_D^*$.
\end{itemize}

\subsection{MARL Algorithms and Frameworks}
This section summarizes the most common algorithms in MARL in Table \ref{tab:marl_algorithms}, focusing on their foundational principles, game-theoretic connections, and unique characteristics tailored to diverse multi-agent scenarios.

\renewcommand{\cellalign}{tl}
\begin{table*}
    \centering
    \caption{MARL Algorithms, Games, and Training Frameworks}
    \ra{1.3}
    % \resizebox{\linewidth}{!}{ % Scale table to column width
    \begin{tabular}{@{}p{2cm}p{7.5cm}p{8cm}@{}}
    \toprule
    \makecell{\textbf{Name /} \\ \textbf{Type of Game}} & \textbf{Algorithm Description} & \textbf{Key Characteristics} \\
    \midrule
    \makecell{\textbf{Q-learning} \\ Stochastic Games} & 
    \raggedright Classical Q-learning adapted for multi-agent environments by treating agents as self-interested players striving to maximize their individual rewards \cite{marl-book}. & 
    \begin{tabular}[t]{@{}p{8cm}@{}}
    \raggedright - Agents are independent learners. \\ 
    \raggedright - Leads by default to non-cooperative dynamics unless coordinated. \\ 
    \raggedright - Often suffers from instability due to non-stationarity caused by simultaneous learning \cite{LITTMAN1994157}.
    \end{tabular} \\
    
    \makecell{\textbf{CTDE} \\ Cooperative Games} & 
    \raggedright Centralized Training with Decentralized Execution (CTDE) is a training framework where agents optimize policies using a centralized critic, while during execution, they make decisions independently based on local observations \cite{foerster2017counterfactualmultiagentpolicygradients}. & 
    \begin{tabular}[t]{@{}p{8cm}@{}}
    \raggedright - Centralized critic enables inter-agent coordination during training, mitigates non-stationarity and provides stable learning signals. \\ 
    \raggedright - Decentralized execution for scalability. \\ 
    \raggedright - One of the most prevalent paradigms in cooperative games \cite{foerster2017counterfactualmultiagentpolicygradients}.
    \end{tabular} \\
    
    \makecell{\textbf{VDN} \\ Fully Cooperative \\Games} & 
    \raggedright A Value Decomposition Network (VDN) decomposes a joint Q-value received by a single global reward function into additive components for individual agents, simplifying learning in cooperative games \cite{rashid2018qmixmonotonicvaluefunction}. & 
    \begin{tabular}[t]{@{}p{8cm}@{}}
    \raggedright - Best for simple cooperative dynamics. \\ 
    \raggedright - Limited in representing complex dependencies and complex inter-agent dynamics \cite{rashid2018qmixmonotonicvaluefunction}.
    \end{tabular} \\
    
    \makecell{\textbf{IPPO} \\ Independent \\Stochastic Games} & 
    \raggedright Independent Proximal Policy Optimization (IPPO) adapts the Proximal Policy Optimization (PPO) algorithm \cite{schulman2017proximalpolicyoptimizationalgorithms}  for independent agents who maximize individual rewards based on local observations, treating them as self-interested entities \cite{dewitt2020independentlearningneedstarcraft}. & 
    \begin{tabular}[t]{@{}p{8cm}@{}}
    \raggedright - Independent policy optimization for each agent. \\ 
    \raggedright - Often results in non-cooperative dynamics and limited coordination. \\ 
    \raggedright - Simple but risks inefficiency in collaborative scenarios \cite{marl-book}.
    \end{tabular} \\
    
    \makecell{\textbf{MAPPO} \\ Cooperative \\Games} & 
    \raggedright Multi-Agent Proximal Policy Optimization (MAPPO) builds on PPO \cite{schulman2017proximalpolicyoptimizationalgorithms}  with a centralized critic to support cooperation among agents during training, while execution remains decentralized \cite{yu2022surprisingeffectivenessppocooperative}. & 
    \begin{tabular}[t]{@{}p{8cm}@{}}
    \raggedright - Centralized critic during training for global value sharing. \\ 
    \raggedright - Balances cooperation during training and scalability in decentralized execution. \\ 
    \raggedright - Effective for cooperative multi-agent tasks \cite{marl-book}.
    \end{tabular} \\
    
    \makecell{\textbf{MADDPG} \\ Mixed-Motive \\Games} & 
    \raggedright Multi-Agent Deep Deterministic Policy Gradient (MADDPG) extends Deep Deterministic Policy Gradient (DDPG) \cite{pmlr-v32-silver14}  for multi-agent settings, enabling learning of both cooperative and competitive dynamics via a centralized Q-function \cite{lowe2020multiagentactorcriticmixedcooperativecompetitive}. This is illustrated in Fig. \ref{fig:decentralized-actor_centralized-critic}. & 
    \begin{tabular}[t]{@{}p{8cm}@{}}
    \raggedright - Centralized critic for joint optimization. \\ 
    \raggedright - Deterministic policies boost stability in continuous action spaces. \\
    \raggedright - The support of cooperative and competitive dynamics, makes it particularly suitable for complex, mixed-motive environments. \\ 
    \raggedright - Stable in continuous action spaces \cite{lowe2020multiagentactorcriticmixedcooperativecompetitive}.
    \end{tabular} \\
    \bottomrule
    \end{tabular}
    % } % End resizebox
    \label{tab:marl_algorithms}
\end{table*}

\begin{figure}[H]
        \centering
        \includegraphics[width=0.45\textwidth, trim=0cm 0cm 0cm 0cm, clip]{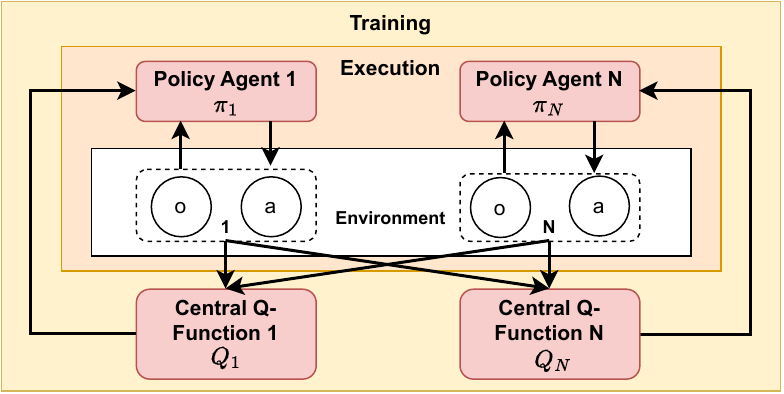} 
        \caption{Centralized training and decentralized execution framework in MADDPG. During training, each agent is associated with its own centralized Q-function ($Q_1$ to $Q_N$), which uses the global state and joint actions of all agents to optimize its policy ($\pi_1$ to $\pi_N$). At execution, each agent acts based on its decentralized policy derived during training, relying only on local observations ($o_1$ to $o_N$) and actions ($a_1$ to $a_N$). This structure effectively allows agents to adapt to competitive, collaborative, and mixed-interest dynamics.}
        \label{fig:decentralized-actor_centralized-critic}
    \end{figure}

Fig. \ref{fig:decentralized-actor_centralized-critic} visualizes the centralized training and decentralized execution framework. During training, agents learn centralized Q-functions, $Q_i(x, a_1, ..., a_N)$, that take as input the joint actions and states of all agents to optimize each agent’s policy, $\pi_i$. This centralized action-value function evaluates joint actions in relation to the global state $x$, which could consist of all agents’ observations $x = (o_1, ..., o_N)$, or additional available state information.

In \emph{competitive dynamics}, each agent optimizes its own centralized Q-function, $Q_i$, in order to optimize its individual reward while considering the joint actions and state $(x, a_1, ..., a_N)$. The reward structure differs for each agent to reflect their opposing objectives.

In \emph{collaborative dynamics}, the centralized Q-function evaluates the joint actions of all agents to maximize a shared global reward, which is uniform across agents, encouraging them to optimize collective performance.

In \emph{mixed-interest dynamics}, the centralized Q-function incorporates individual and shared reward components. This reward components allows agents to balance competing and cooperating incentives. 

\section{Cyber Gyms: Simulation Frameworks}
\label{acas_gyms}
Cyber Gyms are dynamic, interactive environments for the training of MARL agents, addressing the limitations of static datasets that quickly become outdated or biased in rapidly evolving cybersecurity settings. These platforms enable agents to develop robust, cooperative, and adaptive strategies by simulating diverse, unpredictable, and complex attack patterns. Tools like CyberBattleSim \cite{msft:cyberbattlesim} and NASimEmu \cite{janisch2023nasimemunetworkattacksimulator} foster self-play and iterative learning, prepare agents for real-world adversarial scenarios and support sophisticated architectures like Autonomous Intelligent Cyber-defense Agents (AICA).

\subsection{APIs to Train MARL Agents}
MARL APIs allow the training of multiple RL agents in the same environment by considering interactions in cooperative, competitive, and mixed-interest scenarios. Unlike single-agent environments, MARL environments must handle the complexity of agent interactions, including dynamic strategy adaptation, communication, and emergent behaviors.
Standardized MARL environments, such as PettingZoo \cite{terry2021pettingzoogymmultiagentreinforcement}, address these challenges by offering standardized APIs and supporting two different execution paradigms: \emph{parallel execution} and {Agent-Environment Cycle (AEC)} models. 

\emph{Parallel execution} assumes all agents act simultaneously, simplifying environments with large numbers of agents. However, the parallel execution can lead to conflicts such as race conditions. Race conditions occur when the outcome of the environment depends on the order or timing of conflicting actions, which leads to unpredictable or non-deterministic behavior. Tie-breaking can be used to resolve such conflicts, applying rules for conflict resolution or randomness.

In contrast, the AEC model executes its actions sequentially. This prevents race conditions by ensuring that each agent observes an updated state after the previous agent’s action. The AEC model allows the attribution of rewards to specific actions and simplifies debugging. Widely used APIs for the Training of MARL agents are PettingZoo \cite{terry2021pettingzoogymmultiagentreinforcement}, MARLlib \cite{hu2023marllibscalableefficientmultiagent}, PyMARL \cite{samvelyan2019starcraftmultiagentchallenge}, or OpenSpiel \cite{lanctot2020openspielframeworkreinforcementlearning}.

\subsection{Cyber Gyms and Their Role in AICA Development}
Simulations and emulations of real networks are essential for training RL-based autonomous cyber agents. While simulations provide an abstract, fast, and scalable environment ideal for training, they often face a "reality gap," where the simplifications fail to replicate real-world complexities. In contrast, emulations utilize virtual machines and realistic configurations to mirror actual network behaviors but are slower and less scalable due to their resource-intensive nature \cite{janisch2023nasimemunetworkattacksimulator}. Tools like NASimEmu \cite{janisch2023nasimemunetworkattacksimulator} and VINE \cite{10.1145/2808475.2808486} bridge this gap by providing realistic environments for policy evaluation. These environments test architectures like AICA in dynamic threat scenarios, advancing multi-agent reinforcement learning (MARL) integration for proactive and collaborative defenses \cite{janisch2023nasimemunetworkattacksimulator, wilson2024multiagentreinforcementlearningmaritime, finistrella2024multi}.
CyberGyms, such as CyberBattleSim \cite{msft:cyberbattlesim}, play a crucial role in exploring these dynamics. CyberBattleSim models an attacker moving laterally through a network to target specific systems (Capture-The-Flag Exercise). T. Kunz et al. \cite{kunz2023multiagentcyberbattlesimrlcyber} extended this environment with a defense agent, enabling the training of attacker-defender strategies in a multi-agent setting. However, a key challenge remains in applying agents trained in simulated environments, such as CyberBattleSim, to real-world network environments due to the persistent reality gap.
The following sections discuss the cyber gyms identified during the study, addressing their strengths and weaknesses.

One of the best-known environments is \textbf{CyberBattleSim}, developed by Microsoft to model attacks on computer networks. Initially designed as a SARL environment to simulate an attacker moving laterally through a network to compromise specific systems, CyberBattleSim is a powerful showcase to show the power of RL in red-team automation. T. Kunz et al. \cite{ kunz2023multiagentcyberbattlesimrlcyber} extended the framework to include a defense agent, transforming it into a multi-agent setting where attacker-defender strategies can be trained. The framework has its strengths in its simple integration of SARL algorithms and is effective for training basic adversarial actions. Its limitations lie in the abstracted simulations, which lack real-world applicability, and its limited scenarios, which struggle to simulate highly complex setups.

H. Emerson et al. \cite{emerson2024cyborgenhancedgymdevelopment} introduced \textbf{CybORG++}, an environment focusing on multi-agent and adversarial learning in cyber scenarios. Built on the earlier CAGE CybORG framework \cite{baillie2020cyborgautonomouscyberoperations}, CybORG++ supports the training of MARL agents and is designed to model real-world network defense challenges. It enables agents to train adversarial attackers in a red vs. blue team setup. The framework's benefits are its support for multi-agent learning and flexibility for modeling adversarial scenarios. Despite these strengths, CybORG++ faces limitations in scalability for complex, large-scale networks and a simulation reality gap. Notably, CybORG has been integrated into the CAGE 4 Challenge, a MARL-based environment for evaluating autonomous cyber defense agents under partial observability and limited communication \cite{kiely2025exploring}.

\textbf{NASimEmu} \cite{janisch2023nasimemunetworkattacksimulator} allows the simulation of network attacks. It combines simulation and emulation to create realistic cyber-attack scenarios, enabling the training of sophisticated SARL agent policies for cyber attacks. The strength of this environment lies in its high realism for attack-defense scenarios through an emulation option and its support for integrating real-world penetration testing tools. However, a notable downside is that setting up an emulation environment is resource-intensive and requires significant effort to configure and maintain emulated scenarios. Another limitation is that the environment is restricted to SARL agent training.

\textbf{VINE} \cite{10.1145/2808475.2808486} is a robust cyber experimentation environment developed to provide a scalable and configurable platform for studying moving target defenses (MTDs), which are characterized by their ability to adapt and modify system states to enhance security. It allows to create, deploy, and execute complex network scenarios with dynamic background traffic and attack conditions. Designed for high flexibility, VINE facilitates the evaluation of adaptive cyber defense strategies through detailed instrumentation and monitoring.

\subsection{Reference Architecture for AICA}
R. Meier et al. \cite{9467801} and A. Kott et al. \cite{kott2023autonomousintelligentcyberdefenseagent} provide reference architectures for automated and adaptive cyber defense, aimed at enabling NATO to counter sophisticated adversaries in contested battlefields by deploying AICA capable of detecting and neutralizing enemy malware within military tactical networks. The reference architecture incorporates five high-level functions:
\begin{itemize}
    \item \textbf{Sensing and World State Identification:} Enables agents to acquire and process data from their environment, identifying potential risks and triggering appropriate defense mechanisms.
    \item \textbf{Planning and Action Selection:} Allows agents to generate and prioritize response plans, selecting optimal actions based on the situation.
    \item \textbf{Action Execution:} Ensures the implementation and monitoring of defense actions, adjusting strategies dynamically to mitigate threats.
    \item \textbf{Collaboration and Negotiation:} Facilitates cooperation among agents or with centralized command systems, enhancing response coordination.
    \item \textbf{Learning:} Supports continuous improvement of defense strategies through experience and feedback integration.
\end{itemize}
\textbf{Decentralized vs. Centralized Architectures:} The AICA framework supports two different implementation approaches:
\begin{itemize}
    \item \textbf{Centralized Architectures:} These involve a master-agent system where decision-making and data processing are centralized. One possible implementation of such a system can be realized with SARL.
    \item \textbf{Distributed Architectures:} Agents self-organize in a distributed system to detect and defend against cyberattacks, enhancing resilience and fault tolerance, though with added complexity. Such systems can be implemented with MARL.
\end{itemize}

\section{Applications of MARL in Cyber-Defense}
\label{applications}
This section highlights the most significant applications of MARL in current research and its role in addressing critical challenges like intrusion detection, moving target defense, and coordinated threat mitigation.

\subsection{Intrusion Detection and Adaptive Responses}
\begin{figure}%[H]
        \centering
        \includegraphics[width=0.4\textwidth, trim=0cm 0cm 0cm 0cm, clip]{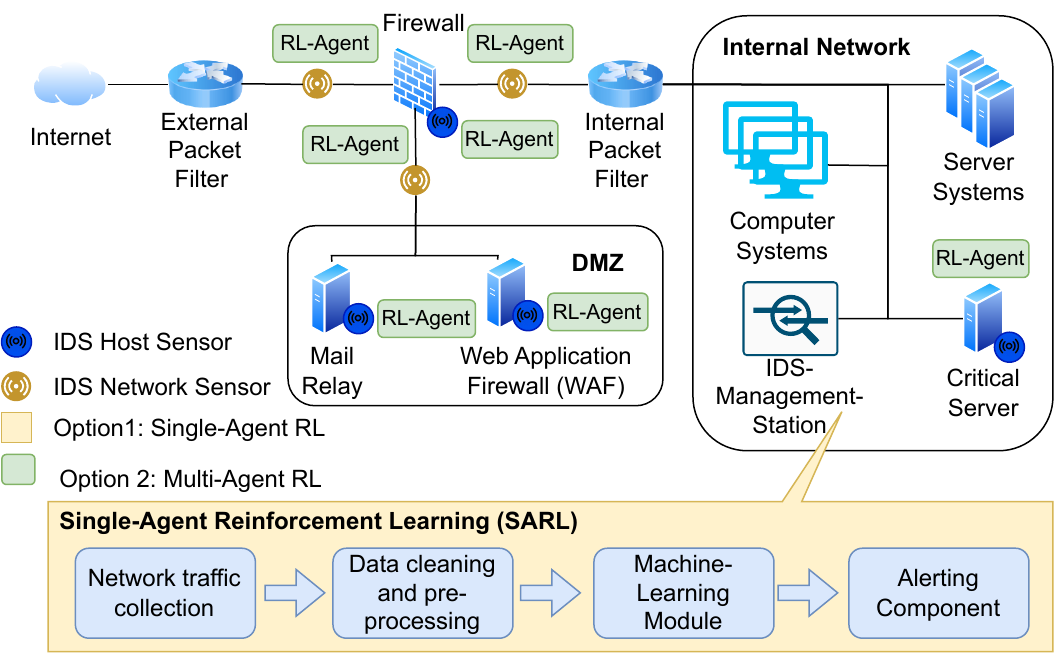} 
        \caption{Comparison of SARL and MARL for Intrusion Detection System (IDS) architectures. In the SARL approach, sensors are deployed at key points (e.g., firewalls, web application firewalls, and critical servers) to collect and analyze network data centrally at an IDS-Management Station managed by a single RL-Agent. In contrast, the MARL approach deploys multiple RL-Agents that collaborate and share information across the network, enhancing decentralized detection of malicious activities.}
        \label{fig:Configuration}
    \end{figure}

RL agents improve intrusion detection by learning adaptive policies that improve detection rates and reduce false positives \cite{yang2024surveydeepreinforcementlearning, alavizadeh2021deepqlearningbasedreinforcement}. However, as highlighted in the surveys \cite{palmer2024deepreinforcementlearningautonomous, finistrella2024multi}, SARL is limited by its inability to effectively manage scalability, dynamic interactions, and decentralized scenarios. MARL overcomes these issues by enabling adaptive and collaborative strategies across multiple agents, which is crucial for handling the complexity and variability of modern cyber threats (Fig. \ref{fig:Configuration}).

\subsection{Red and Blue Team Dynamics}
T. Kunz et al. \cite{kunz2023multiagentcyberbattlesimrlcyber} demonstrated that adversarial training of an attacker (red agent) against a defender (blue agent) enhances the defender's ability to counter sophisticated attacks. This section provides an overview of using RL to train both attack and defense agents. Such an adversarial setup can be formulated as a zero-sum game, emphasizing the importance of training strong attack agents to facilitate the development of equally robust defense agents.

\begin{figure}[H]
        \centering
        \includegraphics[width=0.45\textwidth, trim=0cm 0.35cm 0cm 1.4cm, clip]{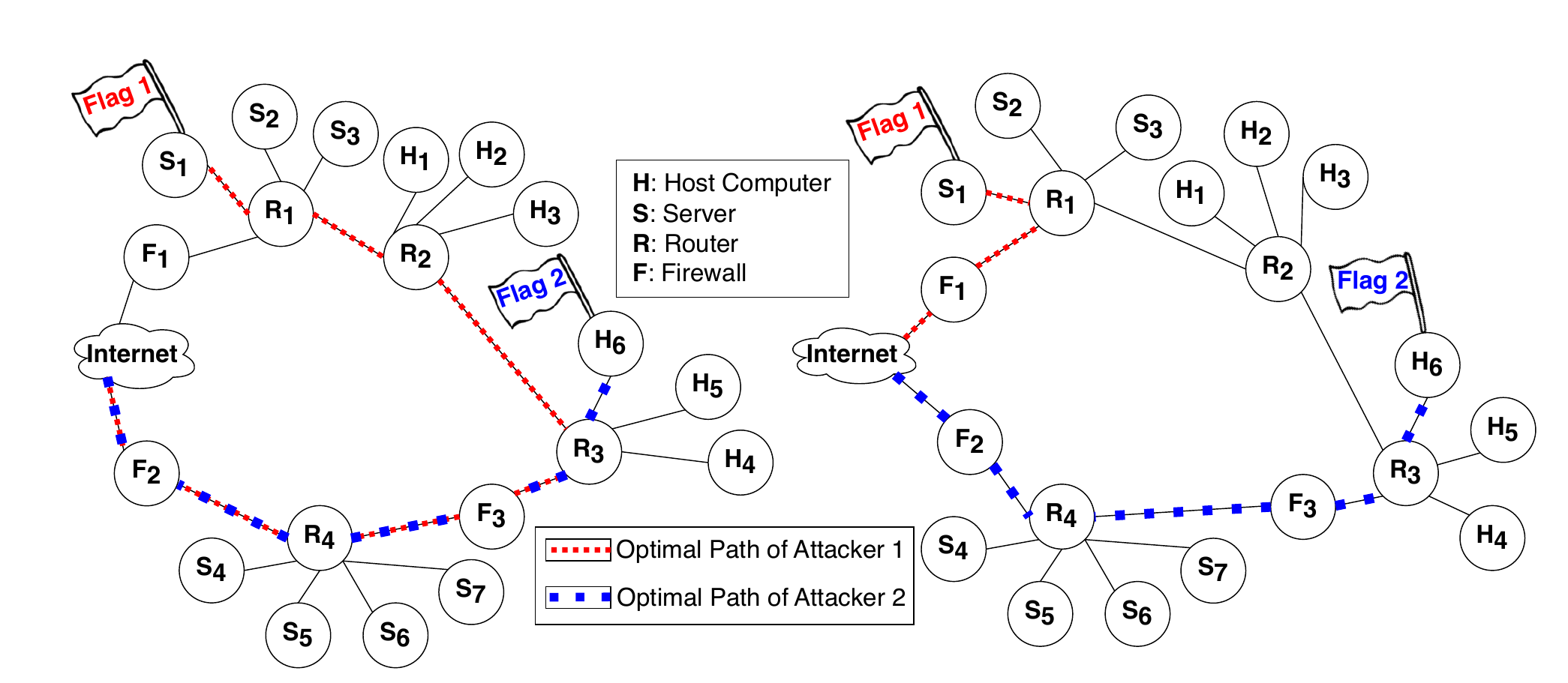} 
        \caption{Multi-attacker scenarios in a network. \textbf{Left:} Attackers 1 and 2 share the same set of exploits but aim for different targets. Both attackers follow their respective optimal paths to achieve their goals. \textbf{Right:} Attacker 1 and Attacker 2 use different sets of exploits and have different targets. This results in divergent optimal paths tailored to their attack capabilities and objectives.}
        \label{fig:Red_Teaming}
    \end{figure}

\paragraph*{Red Teaming} C. Seifert et al. \cite{msft:cyberbattlesim} introduced CyberBattleSim, a platform for training RL-based attackers in abstract enterprise networks with predefined vulnerabilities for lateral movement (Fig. \ref{fig:Red_Teaming}). Attacker performance is evaluated based on the number of steps to achieve network ownership and cumulative rewards over training epochs. This platform is essential for reinforcement learning in red teaming. It provides a controlled, repeatable environment to develop and benchmark advanced attacker strategies, enabling systematic improvement and understanding of RL-based approaches in cybersecurity. Based on this concept, various other environments have been developed for training attack agents with RL, such as NaSimEmu \cite{janisch2023nasimemunetworkattacksimulator}, CYBERSHIELD \cite{10710208}, CSLE \cite{LimmenCSLE} or CybORG++ \cite{emerson2024cyborgenhancedgymdevelopment}.

\begin{figure}[H]
    \centering
    \includegraphics[width=0.4\textwidth, trim=0cm 0.2cm 0cm 0.1cm, clip]{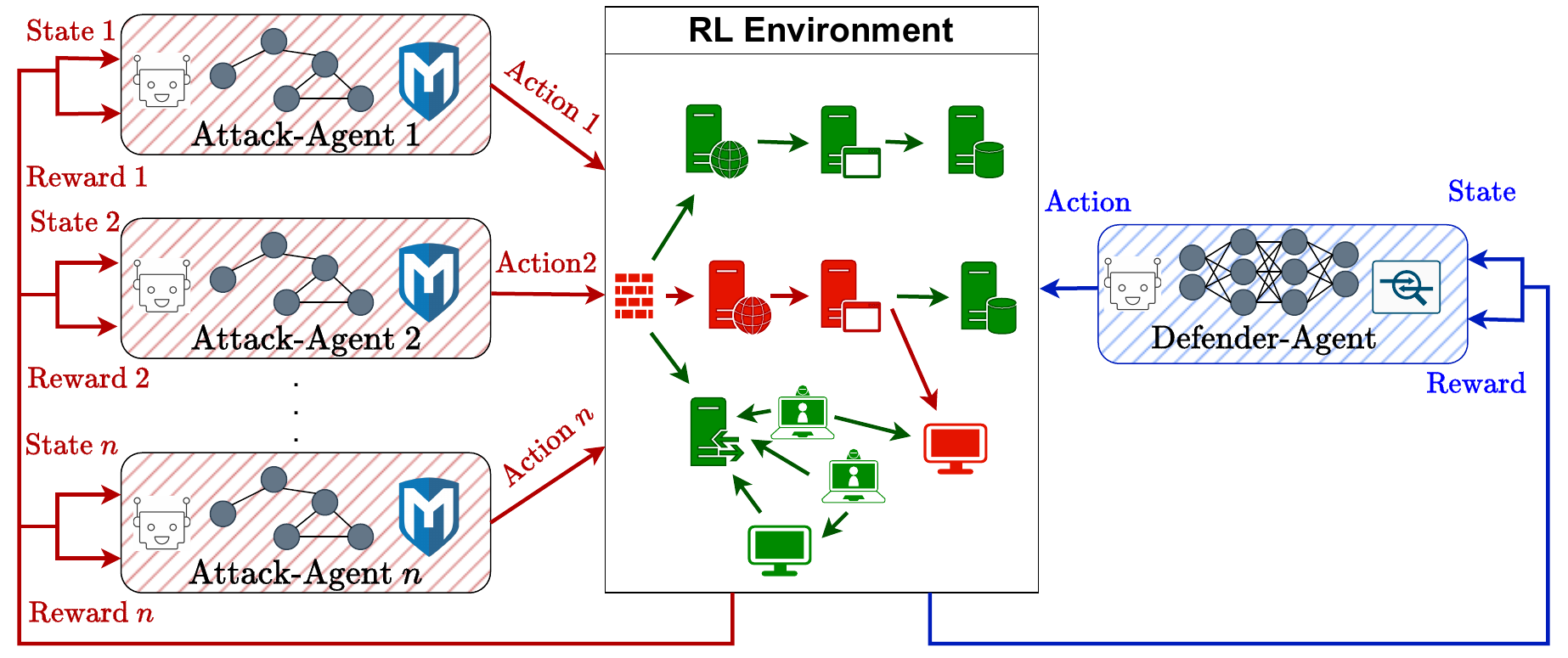} 
    \caption{A swarm of $n$ Attack-Agents collaborates to infiltrate a computer network. The attackers exploit vulnerabilities in the network. This scenario shows a single Defender-Agent, which monitors the entire network, receives global state information and acts to secure the system by preventing or mitigating intrusions.}
    \label{fig:Project_Setup}
\end{figure}

\paragraph*{Blue Teaming} RL defenders counter threats by learning adaptive strategies that dynamically respond to evolving attacks. A. Uprety et al. \cite{Uprety_2021} highlight the ability of RL to secure dynamic IoT environments where real-time adaptation is critical to defend against emerging threats. H. Alavizadeh et al. \cite{alavizadeh2021deepqlearningbasedreinforcement} demonstrate the high accuracy of deep Q-learning in intrusion detection by using autonomous learning to refine defenses against various attacks . S. Jajodia et al. \cite{Hu2019} show how RL can be combined with game theory and moving target defenses (MTD) to disrupt attackers by dynamically changing system configurations. These advances make RL a powerful tool for autonomous and scalable cybersecurity defense. Additionally, T. Kunz et al. \cite{kunz2023multiagentcyberbattlesimrlcyber} emphasize that Multi-Agent Reinforcement Learning (MARL) can significantly enhance blue team effectiveness by training agents in adversarial scenarios, enabling them to counter sophisticated attackers through joint and iterative learning processes, as illustrated in Fig. \ref{fig:Project_Setup}.

\subsection{Lateral Movement Simulation}
\begin{figure}[H]
        \centering
        \includegraphics[width=0.45\textwidth, trim=0cm 0.3cm 0cm 0cm, clip]{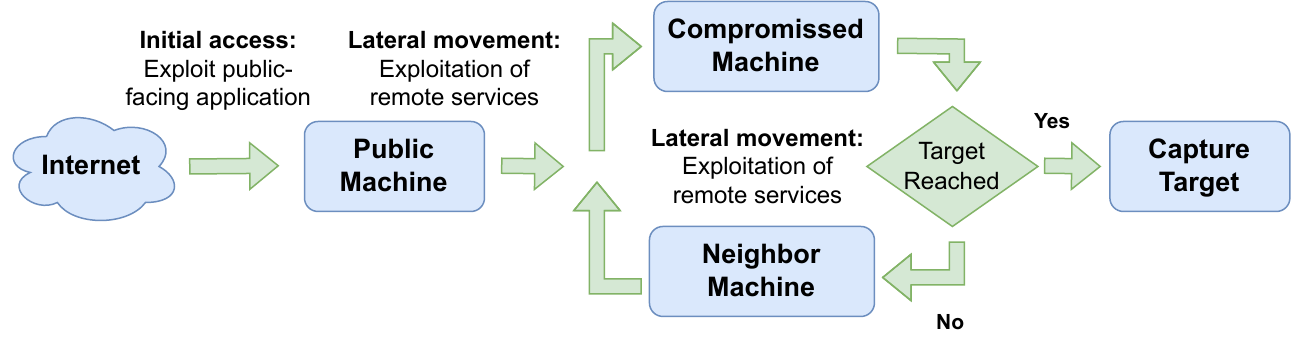} 
        \caption{Attack graph starting from the Internet. The attacker exploits a public-facing application to compromise a public machine. Afterwards, the attacker exploits remote services through lateral movement to neighboring machines. This process continues until the target machine is reached.}
        \label{fig:Attack_graph}
    \end{figure}

In sophisticated, targeted attacks, such as those found in Advanced Persistent Threats (APT), an attacker attempts to move laterally through a computer network to extend their attack to an entire network and ultimately gain control or access to a specific system \cite{Wang_2024}. This process, which is visualized in Fig. \ref{fig:Attack_graph}, usually involves exploiting network vulnerabilities, obtaining additional credentials, and exploiting system connections to extend their reach \cite{mitre_ta0008}. The most important techniques to move laterally through a network include credential dumping and reuse, where attackers extract and use compromised credentials to access connected systems \cite{mitre_t1003}; remote vulnerability exploitation, where vulnerabilities in services such as SMB or HTTP are exploited to penetrate deeper into the network \cite{mitre_t1210}; and privilege escalation, where local vulnerabilities are exploited to gain higher level access and control \cite{mitre_ta0004}. Understanding lateral movement and the interplay between attackers and defenders is essential for developing efficient defense strategies \cite{huang2020farsightedriskmitigationlateral}. Simulation- and training environments such as CyberBattleSim \cite{msft:cyberbattlesim} or NASimEmu \cite{janisch2023nasimemunetworkattacksimulator} provide a controlled environment in which these tactics can be studied. These environments allow researchers to effectively develop and test optimal strategies to counter such threats.

\section{Conclusion}
\label{conclusion}
MARL offers a transformative approach to automating cyber defense by providing adaptive, decentralized, and scalable strategies to counter increasingly sophisticated and dynamic threats. By coordinating agents, MARL offers the technology to create sophisticated AICA and has the potential to significantly improve cybersecurity mechanisms and streamline processes such as penetration testing.

However, despite remarkable progress in the development of MARL algorithms and training methods, practical applications of MARL in cybersecurity still need improvement. Challenges such as managing high-dimensional state and action spaces, ensuring robustness against adversarial attacks, and overcoming deployment gaps between simulations and real-world environments limit MARL's current deployment in cybersecurity. Resolving these issues is critical for MARL to be a valuable tool for advanced cyber defense in practice.

Future research must focus on developing sophisticated simulation and emulation environments, which enable agents to be trained as realistically as possible, further closing the gap between simulation and practice. In addition, integrating agents with generative models such as Generative Adversarial Networks (GANs) could improve scenario diversity and thus enable more effective strategies against complex threats.

By addressing these limitations, MARL can emerge as a critical asset in modern cybersecurity, strengthening distributed systems and networks against increasingly sophisticated and adversarial threats. These advances align with broader efforts to develop intelligent, adaptive cyber defense systems, which are becoming increasingly essential in adversarial environments, including military and defense domains.

\bibliographystyle{IEEEtran}
\bibliography{main}

\end{document}